\documentstyle[sprocl]{article}

\def\Journal#1#2#3#4{{#1} {\bf #2}, #3 (#4)}

\def\PLB{{\em Phys. Lett.}  B}
\def\PRL{\em Phys. Rev. Lett.}
\def\PRD{{\em Phys. Rev.} D}

\def\zz{neutrino }
\def\zzs{neutrinos }

\def\be{\begin{equation}}
\def\ee{\end{equation}}
\def\bea{\begin{eqnarray}}
\def\eea{\end{eqnarray}}

\begin{document}

\title{Looking Back with Neutrinos\footnote {Presented at the
Carolina Symposium on Neutrino Physics, Columbia, March 2000.} }

\author{L. Stodolsky}

\address{Max-Planck-Institut f\"ur Physik, 80805 Munich}


\maketitle

\abstracts{ We briefly discuss the history of suggestions for time-
of-flight effects due to non-zero \zz mass and a recent proposal
that such effects can be used to determine the parameters of
cosmology. With \zzs there is  potentially a much deeper  ``look
back time'' than with photons. We note a new point, that if future
large scale \zz detection arrays see long-time secular variations
in counting rate,  this could be due to highly redshifted bursts
originating in the early universe.}  

\section{Introduction}

Neutrinos are everywhere and they're getting into everything.  Some
have been in the newspapers and  a number of them certainly
penetrated the Pentagon  and have  gone  through secrets at Los
Alamos. Who knows what they won't get into next?

 In view of this interesting if perhaps a bit alarming  situation,
it shouldn't come as a surprise that they're also getting into
history. We all know of course that they play a major role in the
history of the universe. Here, however we want to talk about yet
another role for them: not as  players, but as  reporters.

Back at the time of supernova 1987a and even long  before, it was
realized that if neutrinos had a small mass and if they traveled
over astronomical distances the small deviation of their velocity
from the speed of light could lead to possibly observable effects.
It seems to have begun with Zatsepin \cite{zat}, who in 1968
suggested if a neutrino burst
from a supernova would be detected it could be used to limit the
\zz mass. He used the fact that a mass would mean that
the velocity would be energy dependent. Thus a burst starting out
a few seconds wide in time would, due to the range of energies
contained in it, spread as it traveled to us  to us, and this
spreading would reflect on the mass. He concluded one could improve
the then-existing laboratory mass limit   of 200 eV to  about 2 eV.
Later, with the discovery of different neutrino types came the
realization~\cite{rel} of an even more obvious effect: given
different neutrino types, each one should have a different mass and
so a  different velocity in a burst.  In SN 1987a this lead to a
limit~\cite{raf} of around 10eV for $\nu_e$, unfortunately not very
strong by  present standards.  

 We see that a neutrino mass has two potentially observable 
consequences:

 \indent \indent $\bullet$  A burst spreads  in time due  to the
dispersion of velocities.

 \indent \indent $\bullet$ Different mass states arrive at
different times.

 Note that one of the mass states could  also correspond to photons
emitted in the  event, in which case $m=0$ and the first point
doesn't apply; or contrariwise one of the mass states could be an
as yet undiscovered heavy particle (e.g. a WIMP) in which case the
time-of-flight effects are magnified.  Also note  we should expect
that the \zz mass states are generally  mixed flavor states,
leading to a ``flavor-echo'', as we discussed in ref~[2].

 Then in another vein, there is the possibility of some fundamental
tests of  relativity~\cite{speed}. The limiting velocity for all
forms of radiation should be $c$,  the speed of light. There is no
reason to doubt  this and many good reasons to believe it.
Nevertheless, SN 1987a gave us a rare chance  to test it in a novel
way, and  a rather strong  confirmation it is, on the $10^{-9}$
level.

\section{Measuring the Universe with Neutrinos}
 Here I would like to report on a continuation of this story, which
is again nothing but  kinematics, but nevertheless with amusing
consequences.
Once again 
we are concerned with relativity and neutrino mass. But now it is 
general relativity and instead of learning about the
properties of the \zz we want to use them to determine the large
scale geometry of the universe.

 The main point is the following:  Neutrinos with mass, when
emitted from a  distant source will travel somewhat slower than the
speed of light. This deviation, however, will depend on the
cosmological epoch, since as the universe expands the \zzs slow
down.  When the  neutrino finally reaches us, its total travel time
represents in effect a record of the cosmological epochs it has
passed through.

 This is a simple idea and could easily have been calculated by the
founders  of general relativity and cosmology, if they had known
about the existence of a very light particle which travels almost,
but not quite at the speed of light. But they didn't of course, and
so they concentrated on  ``geometry'' such as the measurement of 
angles subtended by ``standard measuring rods'' or the apparent
brightness of ``standard candles''. Thus it is only very recently, 
stimulated by the growing evidence for neutrino mass~\cite{kam},
that we realized~\cite{me}  there is a ``particle'' as opposed to
a ``geometrical'' way of  surveying the universe.

  Assuming that we know or will know the \zz masses, it turns that
\zz  bursts  from sources with identified red shift can  give us
both  the Hubble constant and the acceleration parameter of
cosmology~\cite{wein} $q$. No independent knowledge of the distance
to the source is necessary, so difficulties involving the ``cosmic
distance ladder'' are absent.

To see how this comes about, let us calculate the difference in
arrival times for two mass states emitted in the same cosmic event. 
We take the standard FRW metric~\cite{wein}  $ds^2=dt^2 -
a^2(t)(d{\bf x})^2$, where we define $a(t)$ to be the expansion
factor of the universe normalized to its present value: $a(t)=
R(t)/R(now)$, so that $a(now)=1$. We proceed by finding  an
equation for the
 coordinate velocity $ dx^i/dt$, where $x^i$ is along the 
particle's flight direction. First we express $ dx^i/dt$ in terms
of   $P^i(t)$,  the  spatial part of the contravariant four-
momentum  $m~dx^{\mu}/ds$.  From the definition
of the metric we have
 $a(t) dx^i/dt =[a(t) P^i(t)]/\sqrt{m^2+[a(t) P^i(t)]^2}$.
 
 Expanding for the relativistic case $P\gg m$ we obtain $a(t)~
dx^i/dt \approx 1-{1\over 2}m^2/[a(t)P^i(t)]^2$
To find $P^i(t)$, we   now make use of the fact that  the covariant
or   ``canonical momentum''
$P_{i}$ is constant (since nothing
depends on the $\bf
x$-coordinate and $P_{i}\sim \partial_i$ ). Furthermore,  since the
different kinds of momenta are  related through the metric tensor,
they all become equal at $t(now)$, where  $a=1$. Hence we can
identify the  constant covariant momentum as $P(now)$, the momentum
at the detector. 
 Thus from  $P^i =g^{ij}P_j=1/(a^2)P_i$, we obtain $P^i
=1/(a^2)P(now)$ . Thus we
finally have
\begin{equation} \label{v}
{dx\over dt}\approx {1\over a(t)}-a(t)
{1\over 2}  {m^2\over P^2(now)}.
\end{equation}
The first term by itself will be recognized as just the equation
for motion along the light cone, and then  there is a small
correction involving the mass.  

 Introducing $\Delta x$ for the difference in the $x$ coordinate of
two different particles of mass
$m_2$ and $m_1$ emitted in the same event at the same time

\begin{equation} \label{deltax}
{d(\Delta x)\over dt}\approx
 a(t){1\over 2} \big[{m_1^2\over
P_1^2(now)}- {m_2^2\over P_2^2(now)}\big] 
\end{equation}

  At the present epoch with $a=1$, $\Delta x$ is just the spatial
separation of the two particles.
Integrating, we have for this separation, or in view of $ v \approx
c=1$ for the time difference in arrival at a detector

\begin{equation} \label{integral}
\Delta t\approx \Delta x\approx \int a(t)~dt {1\over 2}
\big[{m_1^2\over P_1^2(
now)}- {m_2^2\over P_2^2(now)}\big].
\end{equation}

It is thus $\int a(t)~dt$ which ``records''
the cosmological information. Observe $a$ is small at early times
so that most of the effect comes near the present time, as expected
since this is when the \zzs are the ``slowest''. The expressions in
the brackets are the familiar factors giving the difference in
velocity for highly relativistic particles.
 
It now  only remains to get rid of the coordinate dependent $a$ and
$t$ and to re-express things in terms of an observable, namely the
red shift parameter $z$ for the event. With the expansion of $a(t)$
for recent epochs 
$ a(t)= 1+ H
[t-t(now)]-{1\over 2} q  H^2 [t-t(now)]^2+...$,  and the redshift
parameter $z=1/a -1=-H[t-t(now)]+(1+q/2) H^2 [
t-t(now)]^2+...$, we find

\begin{equation}\label{delayz}
 \Delta t \approx {z\over H}\big[1-{3+q\over 2} z +... \big]
{1\over 2} \big[{m_1^2\over P_1^2(
now)}- {m_2^2\over P_2^2(now)}\big]
\end{equation}
 giving the result in
terms of the directly observable $z$.  We thus have the measured
quantities for an event, $\Delta t$ and $z$, given in terms  of  
the present Hubble constant $H$  and the acceleration parameter
$q$. Thus in principle two good events--assuming the neutrino
masses well known by the time this all happens-- fix these
cosmological
parameters. 

This was for the time delay between two distinct mass states.
However it may well be that the mass differences aren't big enough
to give cleanly separated pulses. In that case we can try using the
pulse spreading effect.  So consider the same mass but different
momenta $P$ and $P'$.  We then get a time delay between the two of
\begin{equation} \label{z}
\Delta t \approx {z\over H}\big[1-{3+q\over 2} z +... \big]{1\over
2} m^2 \big[({1\over P(now)})^2- ({1\over P'(now)})^2\big].
\end{equation}

 These formulas are low $z$ expansions, to
order $z^2$.

 The first term in $z$ in the expressions is just what we would get
without general relativity; it says that  $\Delta t$ is simply the
velocity difference times the distance (since $z=H~d$ is just the
Hubble law, where $d$ is the distance). Even this is not entirely
trivial, however,  since it shows how with \zzs one can find H {\it
without knowing the distance to the object}.  That traditional
difficulty of observational cosmology, the ``cosmological distance
ladder'', is gone. It is of course replaced by the difficulty of
detecting \zz bursts at cosmological distances. If the \zz
observations are ever made it will certainly be interesting to
compare the ``\zz H'' found this way   with the ``photon H'' found
by astronomy.

\section{Reality?}

 The realization of these proposals, attractive as they might be,
does not seem immediate or certain. First of all, we need a class
of cosmic events that emits bursts of  \zzs, or \zzs and photons
simultaneously and with great intensity. This does not seem
impossible, and certainly hadronic mechanisms for phenomena like
the gamma ray bursts  would do this, giving neutrinos from charged
pions and photons from  neutral pions.

Then there is the question of detectability. With the bursts coming
from cosmological distances, they will reach us substantially
weakened. Here it seems we must rely on the development of the km
scale detectors in the ocean and in the ice.

     Assuming all this is accomplished we have the question   of
the good separation of the pulses. If our effects   are too small
compared to the duration of the original event itself, our  time of
flight effects will be lost or at least require an elaborate
statistical analysis to be extracted.

To get some feeling for this we can evaluate the kinematic factor
in front of the formulas in terms of an eV (mass)$^2$ for the \zzs
and GeV for their energy:

\begin{equation}\label{delay}
{(m/{\rm eV})^2\over 2 (P/{\rm GeV})^2}\approx   50 \mu {\rm
sec/Mpc}.
\end{equation}

It appears that even at a thousand Mpc, a substantial part of the
way across the visible universe, we may only expect some $\rm msec$
delays. While $\rm msec$ or even $\rm \mu sec$ speed doesn't seem
very difficult for particle detectors, there is the problem of the
intrinsic time scale of the burst itself.  If we take  supernovas
or gamma-ray bursts  as a guide, where the timescale is on the 
order of some seconds, then it seems that to have distinctly
separated bursts we would need to have particles distinctly more
massive than eV's. One possibility might be the third neutrino mass
eigenstate, another perhaps more interesting one would be not the
\zz itself, but a heavy neutral object like the WIMP we are looking
for in our dark matter searches~\cite{dm}. If the objects do indeed
exist  they should be stable and if not overly massive, could be
emitted in high energy bursts. 

 However, there is still the possibility of obtaining information
even if the bursts don't separate into clearly distinguishable
pulses corresponding to the different masses.  There is still the
original Zatsepin effect of the spreading of the 
pulse,  here  represented by Eq~[\ref{z}], and involving  the same
cosmological
information as the  separation effect Eq [\ref{delayz}]. Note our
distinction of the ``separation'' and  ``spreading'' effects is for
the purposes of a qualitative description and  the two may well
overlap, necessitating a detailed analysis including a modeling of
the pulse shapes. Comparing the kinematic factors in
Eq~[\ref{delayz}] and Eq~[\ref{z}]  the condition that the  mass
separation $\Delta t$ be distinctly greater than that of the  pulse
spreading is ${\Delta m^2\over 2m^2}>> {\Delta p\over p}$, where
$\Delta p$ is the energy spread of the burst.

\section{Getting Into the Big Bang}
 Observational methods using the photon, be it the optical photon
of classical  astronomy, the  microwave photon of the  background
radiation or those of radio astronomy, will never allow us to look
further back than
``recombination'',
 some 100,000 years after the Big Bang. One of the most intriguing
potentialities of these \zz or \zz- like based observations   is
that  these particles will  come to us directly from a very  early
epoch.   The ``look-back time'' is  much deeper.
 Decoupling for the \zz   is in the first few minutes,  so with
\zzs  we can, in principle, go back to the first minutes and
  study the space-time geometry at  that time.

 Facing this exciting possibility of looking deep into the Big Bang
are a few requirements and difficulties which are not exactly
trivial. Again, some sort of phenomenon must exist around the epoch
in question in which powerful high energy bursts of \zzs are
emitted. Perhaps collapse of overdense regions to black holes or
annihilation of topological defects are a possibility. Then after
being strongly redshifted and diluted by the expansion of the
universe, the particles must be detected at the earth. Now we no
longer have $z\leq {\cal O}(1)$ but rather, if we go all the way
back to \zz decoupling
 $z\sim (1MeV/0.1meV) \sim 10^{10}$, so that these redshifts and
dilutions will be very great. Furthermore, since the time scale of
the event is  stretched by $z$,  a millisecond event will last a
year upon reaching us. On the other hand this suggests something
interesting: our grandchildren when operating very big \zz arrays
and observing long-time secular variations in the counting rate,
should consider if they are perhaps seeing a burst out of the early
universe and not an instability of their apparatus.

 Alas, in view of the detection difficulties,  this way of doing
cosmology will have to probably
remain fantasy for quite a while, if not forever. On the other
hand, it is hard to think of any  other way  of directly looking at
the pre-recombination epoch. 

In a final mad fantasy we can of course push things even further
back. Imagine that there is the WIMP or some other neutral, very
weakly interacting particle, then we go even further back into the
``Bang''.  Things like the  formation of ``Baby Universes'', or
other Colossal Happenings, presumably involving some activity and
excitations of the particle fields on  a
microscopic timescale, 
 become directly ``visible'' via their bursts, and allow us to
study the space-time structure of the epoch.  While this is indeed
very far-fetched,  it is amusing that at least at the level of
fantasy, there is potentially an observational correspondence to
such happenings.

\end{document}